\documentclass[12pt]{iopart}

\usepackage{graphicx}
\usepackage{dcolumn}
\usepackage{bm}
\usepackage{xcolor}

\newcommand{\Ai}{\mathrm{Ai}}
\newcommand{\RR}{\mathrm{RR}}
\newcommand{\crit}{\mathrm{cr}}
\newcommand{\D}{\mathrm{d}}
\newcommand{\ssb}{\mathbf{s}}
\newcommand{\Eb}{\mathbf{E}}
\newcommand{\Bb}{\mathbf{B}}
\newcommand{\Fb}{\mathbf{F}}
\newcommand{\pb}{\mathbf{p}}
\newcommand{\vb}{\mathbf{v}}
\newcommand{\Omegab}{\bm{\Omega}}

\begin{document}

\title[]{Spin-polarized ${}^3$He shock waves from a solid-gas composite target at high laser intensities}

\author{L Reichwein$^{1,2}$, X F Shen$^1$, M B\"uscher$^{2,3}$, A Pukhov$^1$}

\address{$^1$Institut f\"{u}r Theoretische Physik I, Heinrich-Heine-Universit\"{a}t D\"{u}sseldorf, 40225 D\"{u}sseldorf, Germany}
\address{$^2$Peter Gr\"{u}nberg Institut (PGI-6), Forschungszentrum J\"{u}lich, 52425 J\"{u}lich, Germany}
\address{$^3$Institut f\"{u}r Laser- und Plasmaphysik, Heinrich-Heine-Universit\"{a}t D\"{u}sseldorf, 40225 D\"{u}sseldorf, Germany}

\ead{lars.reichwein@hhu.de}
\vspace{10pt}
\begin{indented}
\item[]December 2023
\end{indented}

\begin{abstract}
We investigate Collisionless Shock Acceleration of spin-polarized ${}^3$He for laser pulses with normalized vector potentials in the range $a_0 = 100-200$. The setup utilized in the 2D-PIC simulations consists of a solid Carbon foil that is placed in front of the main Helium target. The foil is heated by the laser pulse and shields the Helium from the highly oscillating fields. In turn, a shock wave with more homogeneous fields is induced, leading to highly polarized ion beams. We observe that the inclusion of radiation reaction into our simulations leads to a higher beam charge without affecting the polarization degree to a significant extent.  
\end{abstract}

%
\vspace{2pc}

%
\submitto{\PPCF}
%
%
%

\section{Introduction}

In recent years, the acceleration of spin-polarized particle beams from laser-plasma interaction has gained a lot of interest. Applications i.a. include probing the nuclear structure of the proton using deep-inelastic scattering \cite{Glashausser1979}, or polarized fusion, as the cross section for fusion is increased for spin-polarized reactants  \cite{Kulsrud1982}.
	
Most of the recent publications utilize pre-polarized plasma targets, either consisting of HCl \cite{Wu2019, Wu2019a} or $^3$He, which will be considered here. Such a Helium target has recently been experimentally realized and used in a first experiment at PHELIX (GSI Darmstadt) \cite{Zheng2023}. The polarization process is described in detail in the work by Fedorets \textit{et al.} \cite{Fedorets2022}. Pre-polarization is generally required, as no net polarization of the ions can be gained during laser-plasma interaction \cite{Raab2014}. For electron acceleration, an in-situ method of obtaining polarized witness beams in wakefields has been proposed by Nie \textit{et al.} \cite{Nie2021, Nie2022}. A general overview of the state-of-the-art for laser-plasma based acceleration of spin-polarized particle beams is given by Büscher \textit{et al.} \cite{Buescher2020}.
	
The crucial problem with using pre-polarized targets is that the polarization has to be maintained during the acceleration process. However, the strong electromagnetic fields required for high-energy particle beams induce spin precession according to the T-BMT equation \cite{Thomas1926, Bargmann1959}. This was also observed in particle-in-cell (PIC) simulations that study the acceleration of spin-polarized ion beams via Magnetic Vortex Acceleration \cite{Jin2020}. The oscillating laser fields as well as the field structure prevalent in the created plasma channel lead to the depolarization of the accelerated ion beam. Jin \textit{et al.} showed that for a laser with $a_0 = 25$, the final beam polarization is in the range of $82\%$, while for $a_0 = 100$, only $56\%$ could be obtained. Here, $a_0 = e E_0 / (m_e c \omega_0)$ denotes the normalized laser vector potential and $e$ is the elementary charge, $E_0$ the peak laser electric field, $m_e$ the electron rest mass, $c$ the vacuum speed of light and $\omega_0$ the laser frequency.
With the aim of utilizing polarized particle beams in combination with laser intensities in the range of $10^{23}$ W/cm$^2$ \cite{Yoon2021}, or near-future facilities even exceeding that \cite{ELI, XCELS}, different acceleration schemes will be necessary.

 In a proceeding publication \cite{Reichwein2022}, a dual-pulse Magnetic Vortex Acceleration setup was proposed that uses two co-propagating laser pulses with a carrier envelope phase difference of $\pi$. The presence of the two laser pulses creates an accelerating region for ions with decreased depolarization. The setup delivers proton beams with a polarization of $77\%$ even when two pulses with $a_0 = 100$ are utilized. Still, this method will yield decreased polarization for higher intensities as the accelerated particles will still be subject to the strong laser fields.
	
An alternative mechanism avoiding this problem is that of Collisionless Shock Acceleration \cite{Fiuza2012, Daido2012, Zhang2017}. One option to realize this mechanism is to place a solid foil in front of a gaseous target. When the laser pulse irradiates the foil, it heats up the electrons and displaces them with respect to the ions. In turn, an accelerating electric field is induced by which the ions in the gas are being reflected. 
Depending on the laser and target parameters, the laser pulse will not penetrate the foil. Thus, the oscillating laser fields do not reach the polarized ions in the gaseous target, preventing strong depolarization. This scheme of Collisionless Shock Acceleration (CSA) has been proposed for spin-polarized proton beams from an HCl target by Yan \textit{et al.} \cite{Yan2022, Yan2022a}.
In a separate publication, they investigated shock acceleration using a micro-structured foil \cite{Yan2023}.
Alternatively, CSA can also be realized without a foil, utilizing just a density-ramp \cite{Boella2020}.
	
In this paper, we extend the investigation of CSA to regimes of $a_0 = 100 - 200$. In our simulations, we utilize a $^3$He target, whose density is based on current experimental capabilities \cite{Fedorets2022}. We include the effect of radiation damping which becomes important for the electron motion. In turn, the change in electron motion leads to differing electromagnetic fields which affect the ion motion. The setup of our PIC simulations as well as the included spin and radiation effects are explained in section \ref{sec:pic}. The results of the parameter scan concerning laser intensity, foil thickness and the inclusion of radiation reaction are presented in section \ref{sec:results}. The relevance of other effects concerning spin polarization besides precession according to T-BMT is discussed in section \ref{sec:discussion}.

\section{PIC simulation setup} \label{sec:pic}

In the following, we present the simulation setup used for the parameter scans. For all of the 2D simulations, we have used the particle-in-cell code \textsc{vlpl} \cite{Pukhov1999, Pukhov2016}. The simulation box has a size of $120\lambda_L \times  30 \lambda_L$ with a grid size of $h_x = 0.012 \lambda_L$, $h_y = 0.03 \lambda_L$ (with laser wavelength $\lambda_L = 800$ nm). From here on out, $x$ is the direction of laser propagation. As we utilize the rhombi-in-plane Maxwell solver \cite{Pukhov2020}, our time step is required to be $\Delta t = h_x / c$.
	
For all of the simulations, the linearly polarized laser pulse has a duration of $10 \lambda_L / c \approx 27$ fs and a focal spot size of $5\lambda_L = 4$ {\textmu}m (FWHM). At $t = 0$, the center of the laser pulse is located at $x = -20 \lambda_L$. Its normalized laser vector potential $a_0$ will be varied in the range 100-200.

\begin{figure}
    \centering
	\includegraphics[width=0.75\textwidth]{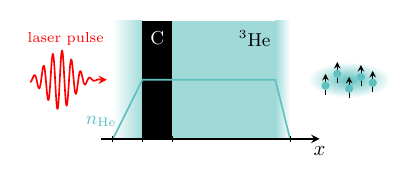}
	\caption{\label{fig:setup2d}Schematic of the simulation setup. The laser pulse irradiates a target consisting of a solid-density Carbon foil and a gaseous, pre-polarized ${}^3$He target.}
\end{figure}

The target the laser pulse interacts with consists of two components, a Carbon foil and a spin-polarized ${}^3$He gas (cf. Fig. \ref{fig:setup2d} for a schematic).  The thickness of the Carbon foil will be varied throughout the parameter scans, but is in the range of multiple laser wavelengths. It is modelled as a rectangular slab with a Carbon ion density of $33.3 n_\crit$. For all simulations, the foil is simulated with 8 particles per cell and -- for reasons of computational efficiency -- fully ionized. 
The polarized Helium is modelled as a homogeneous gaseous target with $2\lambda_L$ of linear up-ramp that is placed as pre-plasma in front of the foil itself. Afterwards, the Helium target has its peak density of $0.02n_\crit$ for $97\lambda_L$ followed by a $20\lambda_L$ down-ramp back to vacuum. Initially, the Helium is fully ionized, and spin-polarized in $y$-direction, i.e. $s_y = 1$ for all ions. Two particles per cell are used for the Helium ions.
	
The spin precession of the macro-particles is calculated according to the T-BMT equation
\begin{equation}
    \frac{\D \ssb}{\D t} = - \Omegab \times \ssb \; , 
\end{equation}
where $\Omegab$ is the precession frequency given as
\begin{equation}
	\Omegab = \frac{q e}{mc} \left[ \Omega_B \Bb - \Omega_v \left( \frac{\vb}{c} \cdot \Bb \right) \frac{\vb}{c} - \Omega_E \frac{\vb}{c} \times \Eb \right] \; .
\end{equation}
The pre-factors in front of the terms containing electric field $\Eb$, magnetic field $\Bb$ and particle velocity $\vb$ are
\begin{equation}
	\Omega_B = a + \frac{1}{\gamma} \; ,  \Omega_v = \frac{a\gamma}{\gamma + 1} \; ,  \Omega_E = a + \frac{1}{\gamma + 1} \; .
\end{equation}
Here, $a$ is the particle's anomalous magnetic moment and $\gamma$ its Lorentz factor. Since the electric field of the shock is pointed in $x$-direction, i.e. the direction of ion acceleration, the term $\vb \times \Eb$ vanishes. Further, magnetic fields are mostly negligible for shock acceleration, meaning that the precession frequency $\Omegab$ will be small. The relevance of other spin-related effects is discussed in section \ref{sec:discussion} as well as the publication by Thomas \textit{et al.} \cite{Thomas2020}.

For the investigation of the effects of radiation on the spin-polarization of the beam we use the following description: the radiation reaction (RR) force is calculated as
	\begin{equation}
		\Fb_\RR = - \frac{2}{3} \alpha \frac{m c^2}{\hbar} \chi^2 G(\chi) \frac{\pb}{\gamma} = - \nu_\RR \pb \; ,
	\end{equation}
	where
	\begin{equation}
		\chi = \frac{e \hbar}{m^2 c^3} \sqrt{\left( \gamma \Eb + \frac{\pb}{m c} \times \Bb \right)^2 - \left( \frac{\pb}{m c} \cdot \Eb \right)^2 } \;
	\end{equation}
    denotes the quantum parameter. The so-called Gaunt factor 
	\begin{equation}
			G(\chi) = - \int_{0}^{\infty} \frac{3 + 1.25 \chi s^{3/2} + 3 \chi^2 s^3}{(1 + \chi s^{3/2})} \Ai' (s) s \; \D s
	\end{equation}
	incorporates the fact that charges will emit less if $\chi \to 1$, i.e. when QED effects become important. As this integral is computationally expensive, we use the approximation
	\begin{equation}
		G(\chi) \approx \left( 1 + 18 \chi + 69 \chi^2 + 73 \chi^3 + 5.806 \chi^4 \right)^{-1/3}
	\end{equation}
	which has been shown to be a reasonable simplification \cite{Esirkepov2015}. For the macro-particles in the PIC simulations, the radiation-corrected momentum is then calculated as 
	\begin{equation}
		\pb_\RR = \frac{\pb_L}{1 + \nu_\RR \Delta t} \; .
	\end{equation}
    This effect will be switched on and off in the following simulations to observe the consequences for the final Helium beam.

    \begin{figure}
        \centering
        \includegraphics[width=\textwidth]{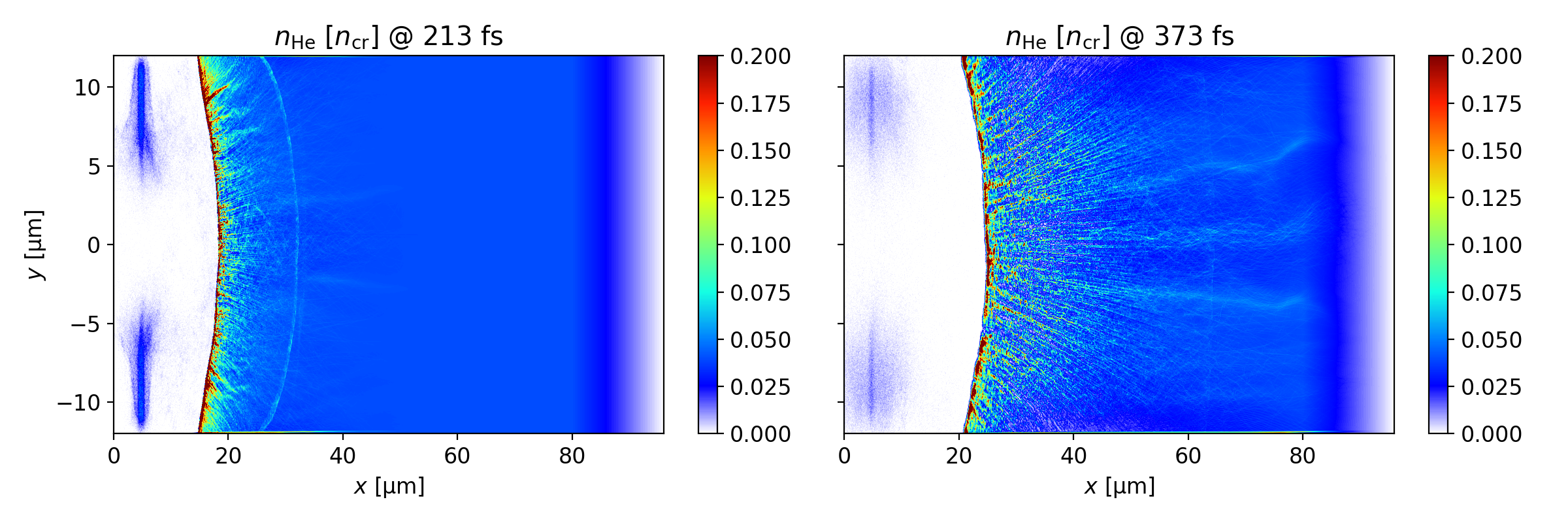}
        \caption{\label{fig:mechanism}Helium density at two different time steps, $t = 213$ fs and $t = 373$ fs, for an $a_0 = 200$ laser pulse and a $2\lambda_L$ thick foil. At earlier times, the shock wave can clearly be identified (high-density region around $x = 20$ {\textmu}m). Later on, a plasma channel similar to the one in the MVA process can be observed, indicating that multiple acceleration mechanisms occur depending on the specific laser and target parameters. The colorbar is clipped at $0.2n_\crit$ for better visibility.}
    \end{figure}

    \section{Results}\label{sec:results}
    We will first discuss the results obtained without considering radiation reaction.
    When the laser pulse irradiates the foil target, it heats up the electrons. The displacement of the electrons with respect to the ions of the foil creates an electric field which reflects the Helium ions. The reflected ions propagate in the form of a shock wave (cf. Fig. \ref{fig:mechanism} at $t = 213$ fs).
    If the foil remains opaque for the laser pulse, the Helium ions only experience the electric field of the shock. This field is much more homogeneous than the oscillating laser fields, thus maintaining a higher spin polarization than in the case of Magnetic Vortex Acceleration, where the laser pulse directly interacts with the target.

    Going to high laser intensities or very thin foils, the foil will be penetrated by the laser pulse. This, in turn, will expose the Helium to parts of the laser pulse itself. This will add further acceleration to ions, but also lower polarization as they now experience the oscillating laser fields. For $a_0 = 200$, we observe signs of a MVA-like plasma channel after $t=373$ fs (cf. Fig. \ref{fig:mechanism}).

    \subsection{Laser intensity}

\begin{figure}
        \centering
        \includegraphics[width=0.75\textwidth]{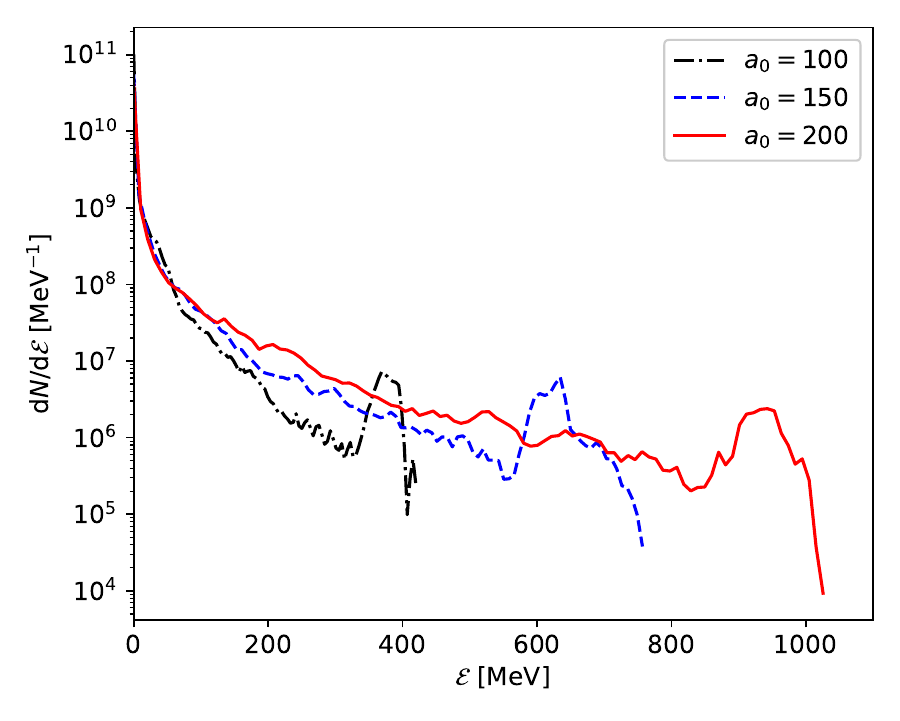}
        \caption{\label{fig:energy_a0}Energy spectrum in dependence of the laser intensity without radiation reaction. For all of the simulations, a foil of $2\lambda_L$ is used. Energy increases with higher laser intensity. A well-defined high-energy peak can is observed for all simulations.}
    \end{figure}

    The influence of the laser intensity on final Helium ion energies is in accordance with previous studies on the CSA process: higher intensity induces a stronger shock potential, i.e. higher energies are achieved. 
    In the case of a foil with $2\lambda_L$ thickness, a well-defined energy peak can be observed in all simulations for the range $a_0 = 100 - 200$ (cf. Fig. \ref{fig:energy_a0}). For $a_0 = 100$, the peak energy is around $\mathcal{E}_p \approx 370$ MeV, while for $a_0 = 200$ up to 943 MeV are obtained (see Table \ref{tab:a0_data}). The scaling found from our simulations is $\mathcal{E}_p \propto a_0^{1.34}$. This scaling is within the theoretical prediction of \cite{Fiuza2012} and stronger than the typical scaling of Target Normal Sheath Acceleration.

    The relative width of the peak (FWHM) drops from 11.5\% to 6.6\% for higher intensities meaning that the condition for particle trapping into the shock wave becomes more restrictive for higher $a_0$. The angular spectrum of the accelerated beam will depend on the laser parameters (duration, focal spot size) as well as the foil thickness. 
    
    The amount of charge accelerated within this peak is on the level of tens of pC. Higher intensities lead to lower beam charge as the shock is less able to trap a significant amount of particles the faster it becomes. For $a_0 = 200$, it drops to about 11.3 pC. This further indicates that our setup is limited by the low Helium density. While choosing a higher density in simulations should improve the amount of accelerated ions, experimentally the Helium-3 target is currently limited to densities of a few $10^{19}\; \mathrm{cm}^{-3}$ \cite{Fedorets2022}. Thus, beam charge can only be further controlled by our choice of laser and foil parameters.

\begin{figure}
        \centering
        \includegraphics[width=0.75\textwidth]{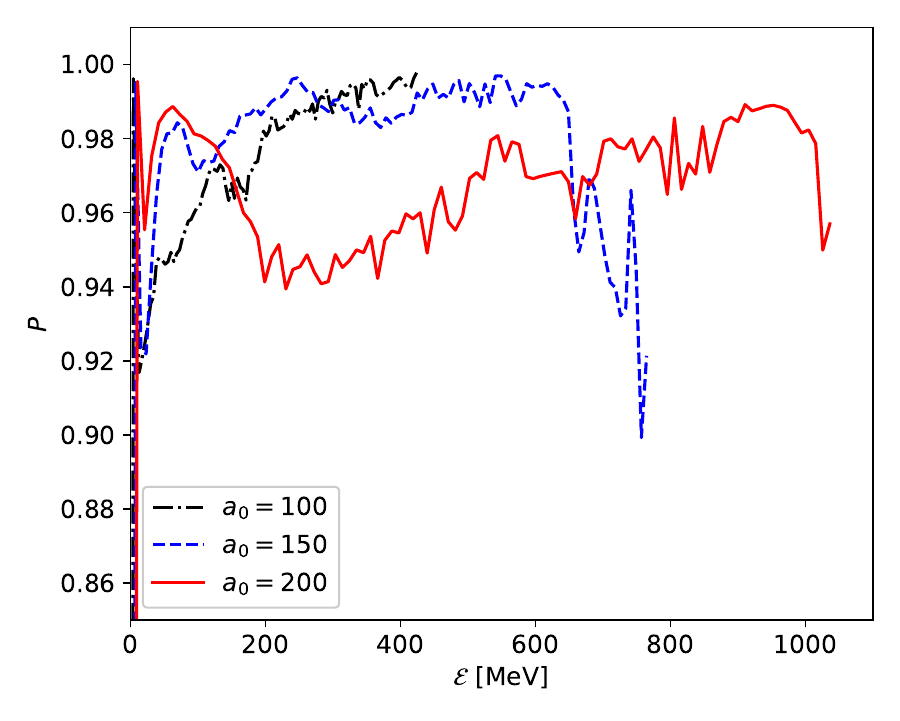}
        \caption{\label{fig:polarization_noRR}Polarization per energy bin for different laser intensities using a $2\lambda_L$ foil without RR. While higher intensity laser pulses will generally lead to lower polarization, the values obtained greatly exceed those observed in MVA simulations.}
    \end{figure}

     The degree of polarization is calculated as $P = \sqrt{P_x^2 + P_y^2 + P_z^2}$, where $P_j = \sum_i s_{i,j} / N$ is the average over the spin components of all $N$ particles in one direction $j\in \lbrace x,y,z\rbrace$. While the minimum degree of polarization per energy bin generally decreases with laser intensity (cf. Fig. \ref{fig:polarization_noRR}), it stays on a 90\% level throughout the $a_0$ scan. Considering only the ions in the FWHM around the energy peak, the degree of polarization is even higher (although the direction of the particle spins may have rotated uniformly compared to their initial direction).
     Once the laser pulse fully penetrates foil, the polarization abruptly drops due to the highly oscillating laser fields. The degree of polarization with the CSA setup is generally much higher than what was obtained in the MVA studies \cite{Jin2020, Reichwein2022}. This is in accordance with the scaling laws derived by Thomas \textit{et al.} \cite{Thomas2020}.

    \begin{table}
    \caption{\label{tab:a0_data}Peak energy, peak width, charge and minimum polarization from simulations with different $a_0$ without and with radiation reaction. For all simulations, the foil was $2\lambda_L$ thick foil. Note that the peak width and the charge correspond to the FWHM around the peak.}
    \begin{indented}
    \item[]\begin{tabular}{lccccc}
        \br
         $a_0$ & RR& $\mathcal{E}_p$ [MeV]& $\Delta\mathcal{E} / \mathcal{E}_p$ [\%] & $Q$ [pC] & $P_\mathrm{min}$ [\%]  \\
         \mr
         100&off& 369.8 & 11.5 & 22.8& 91.65 \\
             125&off& 502.5& 10.1& 25.85& 91.36\\
             150& off&634.6& 8.4& 21.15& 89.39\\
             175&off &791.9& 8.3& 19.9&  89.95\\
             200&off& 942.8 & 6.6& 11.3& 90.98\\
         \mr
          100 & on & 352.9 & 10.2 & 24.25 & 91.49\\
             125 & on & 448.2 & 9.8 &  26.1 & 89.12 \\
             150 & on & 505.4 & 10.3 & 24.75 & 89.49 \\
             175 & on & 603.3 & 10.0 & 23.25 & 84.59 \\
             200 & on & 673.9 & 12.3 & 24.9 & 69.72 \\
         \br
    \end{tabular}
    \end{indented}
    \end{table}

    \subsection{Foil thickness}
    Keeping the laser intensity fixed at $a_0 = 200$, we also perform a scan of the foil thickness. For very thin foils $\leq 1 \lambda_L$, the laser easily penetrates the foil, leading to significantly higher energies but also low beam polarization. In this case, CSA will no longer be the prominent acceleration mechanism at later stages, but rather processes like MVA (as indicated by the similar plasma channel at $t =  373$ fs, cf. Fig. \ref{fig:mechanism}) or, potentially, others. 

\begin{figure}
        \centering
        \includegraphics[width=0.75\textwidth]{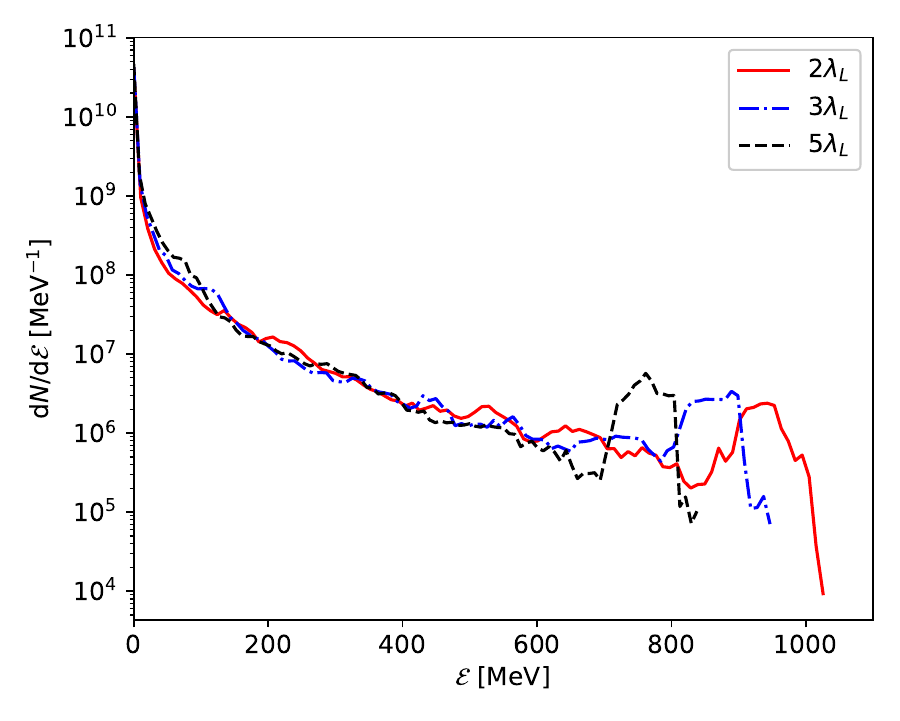}
        \caption{\label{fig:foil_energy}Energy spectra for different foil thicknesses ($a_0 \equiv 200$, RR off). The maximum energy decreases for thicker foils due to the increased shielding of the laser fields and the weaker induced electric shock field. In the case of very thin foils ($\leq 1\lambda_L$), the laser pulse can easily penetrate the foil leading to significantly higher energies, but also highly reduced polarization of the ions.}
    \end{figure}

    For thicker foils in the range $(2-5)\lambda_L$, we see the expected trend of decreasing ion energy: with increasing thickness of the foil, the rear part is heated less sufficiently by the laser pulse, which reduces the shock potential accordingly. At $a_0 = 200$, the original $\mathcal{E}_p \approx 1$ GeV for $2\lambda_L$ drops to 846 MeV for a foil with $5 \lambda_L$ thickness (cf. Fig. \ref{fig:foil_energy}). In all cases, the peak feature is preserved with the relative energy spread staying at a 10\% level. The beam charge is increased from the initial $11.3$ pC at $2\lambda_L$ to $28.7$ pC which is due to the modified shock dynamics. Polarization within the FWHM around $\mathcal{E}_p$ does not change significantly with foil thickness, while the minimum polarization in the whole simulation domain slightly increases due to improved shielding from the laser fields.
    
    \subsection{Radiation reaction}

\begin{figure}
        \centering
        \includegraphics[width=0.75\textwidth]{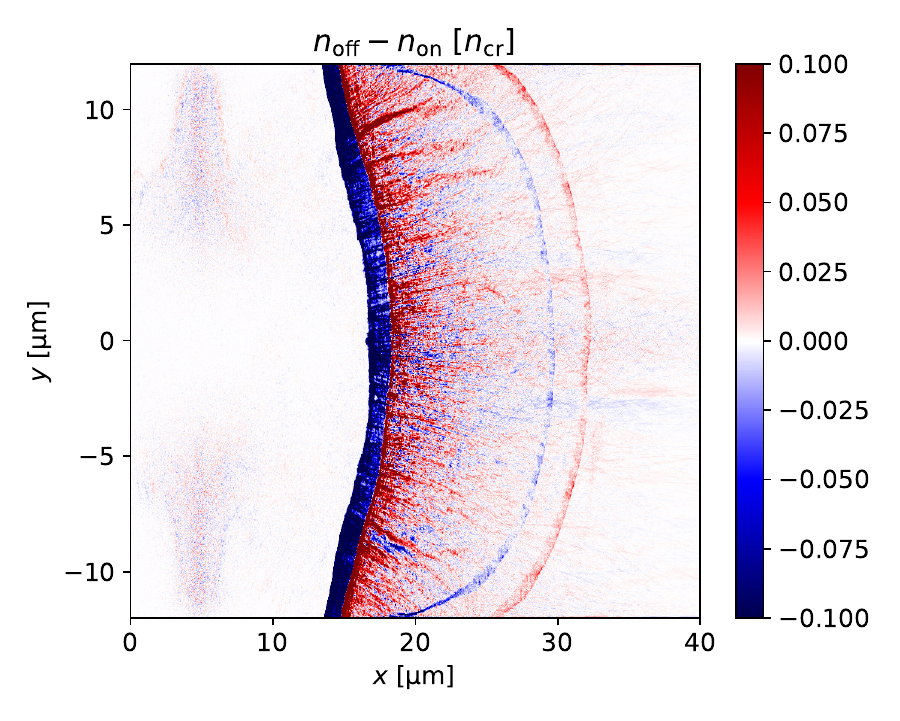}
        \caption{\label{fig:cmp_dens_RR} Comparison of the Helium density after 213 fs for simulations where radiation reaction is turned off or on, respectively. In both cases, the laser had a normalized laser vector potential of $a_0 = 200$ and the foil has a thickness of $2 \lambda_L$. In the case with RR, the shock propagates more slowly (cmp. the red/blue shock waves in the range $x = (10 - 20)$ {\textmu}m as well as the bow-like structures in the region $x = (20 - 30)$ {\textmu}m). The colorbar is clipped at $\pm 0.1 n_\crit$ for better visibility.}
    \end{figure}

    The simulations of the previous sub-sections have not included radiation reaction. In the regime of laser-plasma interaction considered here, however, electrons will radiate part of the energy, thus affecting their motion over the course of the interaction. In particular, the inclusion of RR leads to a slower propagation of the shock wave and therefore lower energy of the accelerated ions. The difference in shock velocity can be seen in Figure \ref{fig:cmp_dens_RR}. 

\begin{figure}
        \centering
        \includegraphics[width=0.75\textwidth]{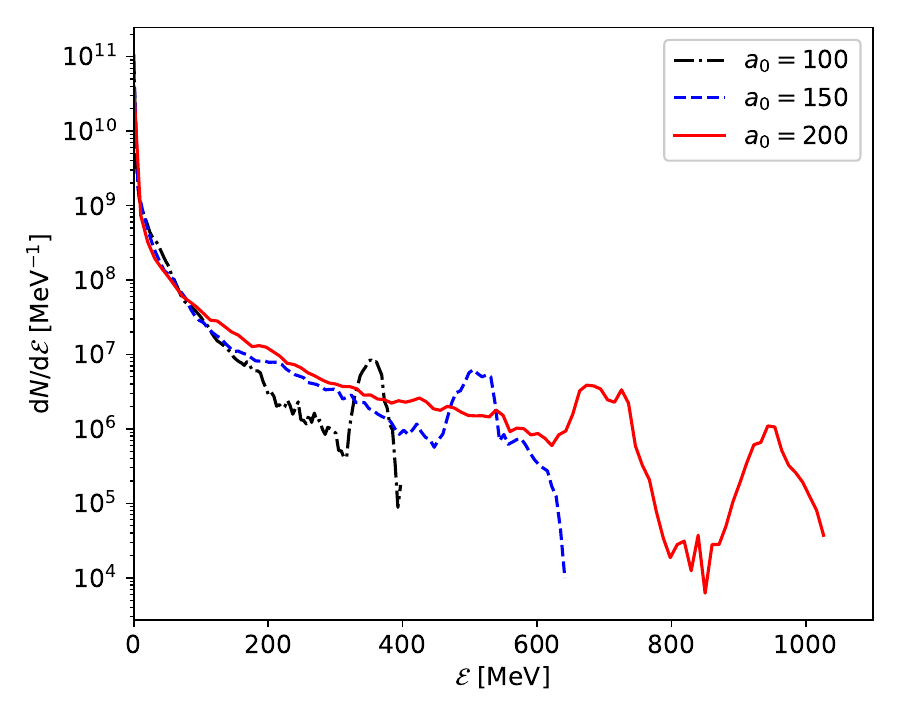}
        \caption{\label{fig:a0_wRR}Energy spectra obtained from simulations with a $2\lambda_L$ foil at different laser intensities under consideration of radiation reaction. While the peak energy generally is reduced compared to the simulations without RR, in the case of $a_0 = 200$ high-energy ions exceeding the peak can be observed, indicating that other acceleration mechanisms besides CSA occur in this regime.}
    \end{figure}

    For $a_0 = 100$, the peak energy drops to approximately 353 MeV, while for $a_0 = 200$ it reaches 674 MeV (see Table \ref{tab:a0_data} for an extensive list of the results). Here it has to be noted, that in the latter case, the shape of the energy spectrum changes significantly: while the main energy peak is moved to lower energies when considering RR, some high-energy ions in the region of the non-RR energy peak remain (see the second peak around 950 MeV in Fig. \ref{fig:a0_wRR}). These ions still experience a fast-moving accelerating field. By comparison, at lower intensities, the shape of the energy spectrum remains the same and only is dampened by RR, i.e. acceleration of the ions is performed solely by the shock wave. The relative width of the main peak stays rather constant at around $10\%$ for all simulations.

    A benefit of radiating damping and the slower moving shock wave is that more particles can be trapped by it: regardless of laser intensity, the beam charge in the FWHM around the energy peak stays consistently at around 25 pC. This leads to the conclusion that a point of saturation with respect to beam charge is reached here, and any further increase would necessitate a change of target parameters (which is, again, experimentally constrained). Polarization of the ion beams in the FWHM remains mostly unchanged when considering RR: while the reduced fields due to RR should generally improve the degree of polarization, it stands to reason that this effect is equilibrated by trapping a larger amount of particles in the shock wave. The minimum polarization throughout the whole simulation box stays at a high degree of approximately $(85-90)\%$ as well, the only exception being the simulation for $a_0 = 200$. Here, the minimum polarization drops to about 70\% in the energy region between the two peaks at 674 MeV and 950 MeV, where significantly fewer particles are located.
    In general, however, we obtain a consistently high degree of polarization with the benefit of accelerating more charges when considering radiation reaction.

    \section{Discussion} \label{sec:discussion}
    Throughout the conducted simulations, the only spin-related effect considered has been precession according to the T-BMT equation. Other effects that would -- depending on the parameter regime -- need to be considered are those of the Stern-Gerlach force \cite{Gerlach1922} and radiative polarization as in the Sokolov-Ternov effect \cite{Ternov1995}. Taking the scaling laws derived in the publication by Thomas \textit{et al.} \cite{Thomas2020}, we can conclude that the Stern-Gerlach force on the Helium ions can be neglected as the differences in particle trajectories would be miniscule.
    For electrons, it is well known from several publications that in the high-intensity regime, their polarization will significantly be affected by radiative processes \cite{Li2019, Guo2020}. Technically, the spin-dependence of the Gaunt factor would also need to be considered in that case \cite{Seipt2023}. In the current publication, however, electron polarization is not of interest and their trajectories will not be significantly altered. The Helium ions themselves will not radiate to a large extent, as their Lorentz factor $\gamma = \mathcal{E} / \mathcal{E}_0 + 1$ is comparatively low even for $a_0 = 200$. Thus, polarization build-up due to radiation emission can be neglected for the Helium ions.

    Another effect that can be neglected here is that of electron-positron pair creation as the laser vector potential never exceeds $a_0=200$ in our simulations. It should, however, be noted that in the regime where pair production can occur, the pre-plasma in front of the foil will likely play an important role for the acceleration process: Wang \textit{et al.} showed that for ultra-intense laser pulses QED cascades in the pre-plasma can lead to the formation of an opaque particle layer in front of the foil \cite{Wang2017}. This, in turn, changes the interaction of the laser pulse with the foil and the subsequent particle acceleration.

    The simulations conducted here are currently restricted to a 2D geometry for reasons of computational efficiency. As it is known from several publications, this means that the beam energy of particles will be overestimated compared to 3D-PIC simulations \cite{Sgattoni2012, Stark2017}. As the results already have indicated, which acceleration mechanism is prevalent during laser-solid ineraction (and in some circumstances, subsequent laser-Helium interaction) will strongly depend on choice of laser and target parameters. Thus, a future, separate study will further investigate the different acceleration mechanisms during high-intensity laser-solid interaction and their consequences for beam polarization. This will include a study of the influence of laser polarization on the acceleration mechanisms. It is e.g. known from the works by Tamburini \textit{et al.} that for Radiation Pressure Acceleration, circular polarized laser pulses exhibit less spatial anisotropies and no significant RR effects compared to linear polarization \cite{Tamburini2012}.

    \section{Conclusion}
    We have studied the acceleration of spin-polarized Helium-3 via Collisionless Shock Acceleration for high laser intensities. A solid Carbon foil is placed in front of near-critical, pre-polarized Helium to shield it from the oscillating laser fields. The obtained particle beams are of significantly higher polarization than what was obtained for previous Magnetic Vortex Acceleration studies. Depending on the laser and target parameters, the foil will be penetrated laser pulse, leading to additional acceleration mechanisms and reduced beam polarization.
    When radiation reaction is considered, the shock wave slows down due to the reduced shock potential, thus allowing a larger amount of Helium ions to be accelerated by it while maintaining a high degree of spin polarization.
    Future research could investigate the possibility of utilizing density ramps to induce similar shock waves in polarized gaseous targets without the need for the Carbon foil. Going to even higher intensities for future laser facilities, the inclusion of pair production and its effects on pre-plasma opaqueness would be of interest for CSA.

    \ack
    This work has been funded in parts by the DFG (project PU 213/9-1). The authors gratefully acknowledge the Gauss Centre for Supercomputing e.V. \cite{gcs} for funding this project (qed20) by providing computing time through the John von Neumann Institute for Computing (NIC) on the GCS Supercomputer JUWELS at Jülich Supercomputing Centre (JSC). The work of M.B. has been carried out in the framework of the JuSPARC (J\"{u}lich Short-Pulse Particle and Radiation Center \cite{jusparc}) and has been supported by the ATHENA (Accelerator Technology Helmholtz Infrastructure) consortium. 
	L.R. would like to thank S. Villalba-Ch\'{a}vez and M. Filipovic for helpful discussions throughout the project.


\providecommand{\newblock}{}

\end{document}